\documentclass[12pt,a4paper]{article}
\usepackage{amsfonts,latexsym}
\usepackage{amsmath,amssymb}
\usepackage{graphicx,color}
\usepackage{multirow}
\numberwithin{equation}{section} \oddsidemargin 0 mm \evensidemargin
0 mm \topmargin -10 mm \textheight 215 mm \textwidth 163 mm

\renewcommand{\thefootnote}{\fnsymbol{footnote}}

\begin{document}
\vspace{12mm}

\begin{center}
{{{\Large {\bf Rank-3 finite temperature logarithmic conformal field theory}}}}\\[10mm]

{Taeyoon Moon$^{a}$\footnote{e-mail address: tymoon@sogang.ac.kr}
and  Yun Soo Myung$^{b}$\footnote{e-mail address:
ysmyung@inje.ac.kr},
}\\[8mm]

{{${}^{a}$ Center for Quantum Space-time, Sogang University, Seoul, 121-742, Korea\\[0pt]
${}^{b}$ Institute of Basic Sciences and School of Computer Aided Science, Inje University Gimhae 621-749, Korea}\\[0pt]
}
\end{center}
\vspace{2mm}

\begin{abstract}
We construct a rank-3 finite temperature logarithmic conformal field
theory (LCFT)  starting from a higher-derivative scalar field model
in the BTZ black hole background. Its zero temperature limit reduces
to  a rank-3 LCFT in the AdS$_3$ background. For a tricritical
generalized massive gravity, we read off  the log-square quasinormal
frequencies of graviton from the poles of the retarded Green's
function in the momentum space. After using the truncation process,
we find quasinormal frequencies from a unitary conformal field
theory. Finally,  employing the  retarded Green's functions on the
boundary, we compute the absorption cross sections of BTZ black hole
which show feature of higher-order differential equations for
scalars.
\end{abstract}
\vspace{5mm}

{\footnotesize ~~~~PACS numbers: }

\vspace{1.5cm}

\hspace{11.5cm}{Typeset Using \LaTeX}
\newpage
\renewcommand{\thefootnote}{\arabic{footnote}}
\setcounter{footnote}{0}

\section{Introduction}
Recently, critical gravities have  been an actively interesting
subject because they were considered as toy models for quantum
gravity~\cite{Li:2008dq,Lu:2011zk,Deser:2011xc,Porrati:2011ku}. At
the critical point of avoiding the ghosts, a degeneracy takes place
and massive gravitons coincide with massless gravitons. Instead of
massive gravitons, an equal amount of logarithmic modes is
considered  at the critical point. According to the AdS/LCFT
correspondence, one finds that a rank-2 logarithmic conformal field
theory (LCFT) is dual to a critical
gravity~\cite{Grumiller:2008qz,Myung:2008dm,Maloney:2009ck}.
However, one has to resolve  the non-unitarity issue of these
log-gravity theories.  In order to resolve the non-unitarity issue,
a polycritical gravity was  introduced to provide multiple critical
points~\cite{Nutma:2012ss} whose CFT dual is  a higher-rank LCFT. On
later, it was shown that a consistent unitary truncation of
polycritical gravity may be realized  at the linearized level for
odd rank~\cite{Kleinschmidt:2012rs}.

The rank of the LCFT refers to the dimensionality of the Jordan
cell.  Explicitly, the LCFT dual to critical gravity has rank-2 and
thus, an operator has its logarithmic partner.  The LCFT dual to
tricritical gravity has rank-3 and  an operator has two logarithmic
partners. Most of all, a truncation allows an odd-rank LCFT to be a
unitary conformal field theory (CFT)~\cite{Bergshoeff:2012sc}. A
six-derivative gravity in three dimensions was treated as dual to a
rank-3 LCFT~\cite{Bergshoeff:2012ev}, while a four-derivative
critical gravity in four dimensions was considered as dual to a
rank-3 LCFT~\cite{Johansson:2012fs}.

One can construct a rank-3 parity odd theory in the context of
four-derivative gravity known as three-dimensional generalized
massive gravity (GMG)~\cite{LS}.  There exist two tricritical points
in the GMG parameter space whose dual theory is a rank-3
LCFT~\cite{Grumiller:2010tj}. At this stage,  we would like to point
out  the following difference in AdS/LCFT correspondence:

\indent $\bullet$ tricritical GMG on the AdS$_3 \leftrightarrow$
 a rank-3 (zero temperature) LCFT

\indent $\bullet$ tricritical GMG on the BTZ black
hole~\cite{Banados:1992wn} $ \leftrightarrow$ a rank-3 finite
temperature~LCFT. \\
Around the BTZ black hole, the authors~\cite{KMP} have confirmed the
AdS/LCFT correspondence between the tricritical GMG and a rank-3
finite temperature LCFT by computing quasinormal frequencies of
graviton approximately. In this computation, one could not construct
a rank-3 finite temperature LCFT whose zero temperature limits
correspond to (\ref{ac1})-(\ref{ac4}).  Therefore, one could not
make a truncation process to obtain the quasinormal modes  from
simple poles of CFT.

In this work, we will construct a rank-3 finite temperature LCFT
starting from a higher-derivative scalar field model in the BTZ
black hole background.  Its zero temperature limit reduces nicely to
a rank-3 LCFT (\ref{ac1})-(\ref{ac4}) in the AdS$_3$ background.  We
obtain the retarded Green's functions from two-point functions. For
a tricritical GMG, we read off the log-square quasinormal
frequencies of graviton from the poles of the retarded Green's
function in the momentum space. Implementing the truncation process,
we find quasinormal frequencies from a unitary finite-temperature
CFT. Finally, employing the retarded Green's functions on the
boundary, we compute the absorption cross sections of BTZ black hole
which involve feature of higher-order differential equations for
scalars.

\section{Rank-3 LCFT on the BTZ black hole}
Let us first consider a gravity action with three scalar fields
$\Phi_1,~\Phi_2,$ and $\Phi_3$ in three
dimensions~\cite{Bergshoeff:2012sc}
\begin{eqnarray}\label{maina}
S=\int d^3x
\sqrt{-g}\Big[R-2\Lambda-\partial_{\mu}\Phi_1\partial^{\mu}\Phi_3
-\frac{1}{2}(\partial_{\mu}\Phi_2)^2-\Phi_1\Phi_2-m^2\Phi_1\Phi_3
-\frac{1}{2}m^2\Phi_2^2\Big].
\end{eqnarray}
Varying for the fields $g^{\mu\nu}$, $\Phi_{3}$, $\Phi_2$, and
$\Phi_1$ on the action \eqref{maina} leads to the equations of
motion,
\begin{eqnarray}
\delta g^{\mu\nu};~~~R_{\mu\nu}-\frac{1}{2}g_{\mu\nu}R+\Lambda
g_{\mu\nu}&=&T_{\mu\nu}
\label{eeq0}\\
\delta \Phi_3;\hspace*{4.35em}(\nabla^2-m^2)\Phi_1&=&0,\label{eeq1}\\
\delta \Phi_2;\hspace*{4.35em}(\nabla^2-m^2)\Phi_2&=&\Phi_1\label{eeq2},\\
\delta
\Phi_1;\hspace*{4.35em}(\nabla^2-m^2)\Phi_3&=&\Phi_2\label{eeq3},
\end{eqnarray}
where the energy momentum tensor $T_{\mu\nu}$ is given by
\begin{eqnarray}
T_{\mu\nu}&=&\partial_{\mu}\Phi_1\partial_{\nu}\Phi_3
+\frac{1}{2}\partial_{\mu}\Phi_2\partial_{\nu}\Phi_2
\nonumber\\
&&\hspace*{2em}-\frac{1}{2}g_{\mu\nu}\Big(
\partial_{\rho}\Phi_1\partial^{\rho}\Phi_3
+\frac{1}{2}(\partial_{\rho}\Phi_2)^2+\Phi_1\Phi_2+m^2\Phi_1\Phi_3
+\frac{1}{2}m^2\Phi_2^2\Big).\nonumber
\end{eqnarray}
We note that varying the action with respect to $\Phi_3(\Phi_1)$
leads to the equation for $\Phi_1(\Phi_3)$, respectively. This
feature will persist in deriving dual CFT operators.

 In this work, we introduce a background metric $\bar{g}_{\mu\nu}$ of BTZ black hole with $\Lambda=-1/\ell^2=-1$
\cite{Banados:1992wn},
\begin{eqnarray}\label{btzm}
ds^2_{\rm B}&=&\bar{g}_{\mu\nu}dx^{\mu}dx^{\nu}\nonumber\\
&=&-\frac{(r^2-r_+^2)(r^2-r_-^2)}{r^2}dt^2+
\frac{r^2}{(r^2-r_+^2)(r^2-r_-^2)}dr^2
+r^2\left(d\phi+\frac{r_+r_-}{r^2}dt\right)^2,
\end{eqnarray}
being a solution to Eqs. (\ref{eeq0})-(\ref{eeq3})
together with
\begin{equation}
\bar\Phi_{i}=0 ~~{\rm  with}~~ i=1,2,3. \end{equation} Here, the ADM
mass $(M)$, angular momentum $(J)$, and right/left temperature
$(T_{R/L})$ are given by \begin{equation} M=r_+^2-r_-^2,~J=2r_+r_-,~
T_{R/L}=\frac{r_+\pm r_-}{2\pi}. \end{equation}

We consider  the perturbation
\begin{equation} g_{\mu\nu}=\bar{g}_{\mu\nu}+h_{\mu\nu},~~
\Phi_{i}=\bar\Phi_{i}+A_{i} \end{equation} around the  BTZ black
hole spacetimes. The metric perturbation $h_{\mu\nu}$ has no
physical degrees of freedom in three dimensions and thus, we will
not consider this further. The relevant quantities are scalar
perturbations whose linearized equations are given by
\begin{eqnarray}
(\nabla_{B}^2-m^2)A_1&=&0,\label{eq1}\\
(\nabla_{B}^2-m^2)A_2&=&A_1\label{eq2},\\
(\nabla_{B}^2-m^2)A_3&=&A_2\label{eq3},
\end{eqnarray}
which lead to the higher-order equations for $A_2$ and $A_3$ as
\begin{eqnarray} \label{aa2}
(\nabla_{B}^2-m^2)^2A_2&=&0, \\
\label{aa3}(\nabla_{B}^2-m^2)^3A_3&=&0.
\end{eqnarray}
This means that two scalars $A_1$ and $A_2$ are auxiliary fields
which are introduced to lower the number of derivatives in the
bilinear action $\delta S_{\rm bilinear}(h_{\mu\nu},A_i)$. One
relevant field is $A_3$ which satisfies the six-derivative equation
(\ref{aa3}). This perturbative picture for scalar fields $A_i$ is
equivalent to ignoring the back reaction of the scalars on the
metric~\cite{Bergshoeff:2012sc}.

We wish to compute the two-point correlation functions on the
boundary and show that they form a rank-3 finite temperature LCFT.
The scalars $A_i$ are written in terms of bulk-to-boundary
propagators $K_{ij}$ with $i,j=1,2,3$, which relate the bulk
solution to the boundary fields $A_{i(b)}$.
 There are two
possibilities~\cite{Kogan:1999bn} to choose the propagators
$K_{ij}$: The first one is the symmetric case where  the propagators
$K_{ij}$ must be symmetric ($K_{ij}=K_{ji}$) and a bulk-to-boundary
propagator  for a massive scalar is chosen to be a non-diagonal
propagator~\cite{Myung:1999nd}. The second one is the Jordan type
(or non-symmetric case) \cite{Bergshoeff:2012sc} where $K_{ij}$ is
given by
\begin{equation}
K_{ij}=\left(
  \begin{array}{ccc}
    K_1 & 0 & 0\\
    K_2 & K_1 & 0 \\
    K_3 & K_2 & K_1\\
  \end{array}
\right) \label{mhme}.
\end{equation}
Here $K_{1}~(=K_{11}=K_{22}=K_{33}),~K_2~(=K_{21}=K_{32})$, and
$K_3~(=K_{31})$ satisfy the following relation:
\begin{eqnarray}
(\nabla^2_{\rm B}-m^2)K_1&=&0,\label{keq1}\\
(\nabla^2_{\rm B}-m^2)K_2&=&K_1\label{keq2},\\
(\nabla^2_{\rm B}-m^2)K_3&=&K_2\label{keq3}.
\end{eqnarray}
In this case, $K_{1},~K_2,$ and $K_3$ correspond to the
bulk-to-boundary propagators of the Klein-Gordon  mode (1), log-mode
(2), log$^2$-mode (3), respectively for the AdS$_3$ background.

Upon choosing the Jordan type,   the solutions to the
Eqs.(\ref{eq1})-(\ref{eq3}) are given by
\begin{eqnarray}
\hspace*{-2.5em}A_1(r,u_+,u_-)&=&\int du_+^{\prime}du_-^{\prime}
\Big[K_{11}(r,u_+,u_-;u_+^{\prime},u_-^{\prime})A_{1(b)}\Big]\nonumber\\
\hspace*{-2.5em}A_2(r,u_+,u_-)&=&\int du_+^{\prime}du_-^{\prime}
\Big[K_{22}(r,u_+,u_-;u_+^{\prime},u_-^{\prime})A_{2(b)}
+K_{21}(r,u_+,u_-;u_+^{\prime},u_-^{\prime})A_{1(b)}\Big]\nonumber\\
\hspace*{-2.5em}A_3(r,u_+,u_-)&=&\int du_+^{\prime}du_-^{\prime}
\Big[K_{33}(r,u_+,u_-;u_+^{\prime},u_-^{\prime})A_{3(b)}
+K_{32}(r,u_+,u_-;u_+^{\prime},u_-^{\prime})A_{2(b)}\nonumber\\
\label{a3}&&\hspace*{14.5em}
+K_{31}(r,u_+,u_-;u_+^{\prime},u_-^{\prime})A_{1(b)}\Big],
\end{eqnarray}
where $u_\pm=\phi\pm t,$ and $A_{i(b)}$ are functions of
($u_+^{\prime},u_-^{\prime}$).

 It is well known that in the BTZ black hole background,
$K_1$ can be found as the solution to a homogeneous Klein-Gordon
equation (\ref{keq1})~\cite{KeskiVakkuri:1998nw}
\begin{eqnarray}
&&K_1(r,u_+,u_-;u_+^{\prime},u_-^{\prime})\nonumber\\
&&=N\left[\frac{\pi^2 T_R T_L}{\frac{r_+^2-r_-^2}{4r}e^{(\pi
T_L\delta u_++\pi T_R\delta u_-)}+r\sinh(\pi T_L\delta u_+)\sinh(\pi
T_R\delta
u_-)}\right]^{\triangle}\label{k1eq}\\
&&\equiv
N\Big[f(r,u_+,u_-;u_+^{\prime},u_-^{\prime})\Big]^{\triangle}\nonumber,
\end{eqnarray}
where $\delta u_{\pm}=u_{\pm}-u_{\pm}^{\prime}$,
$\triangle(\triangle-2)=m^2$, and the Hawking temperature $T_H$ is
defined by $T_{H}=2/(1/T_R +1/T_L)$. Here $N$ is a normalization
constant to be fixed.  Using two relations
\begin{eqnarray}
K_2&=&\frac{\partial K_1}{\partial
m^2}=\frac{1}{2(\triangle-1)}\frac{\partial K_1}{\partial\triangle},\\
K_3&=&\frac{1}{2}\frac{\partial K_2}{\partial
m^2}=\frac{1}{4(\triangle-1)}\frac{\partial K_2}{\partial\triangle},
\end{eqnarray}
$K_2$ and $K_3$ can be expressed  in terms of $K_1$ as
\begin{eqnarray}
K_2&=&\frac{K_1}{2(\triangle-1)}\Bigg(\ln[f]+\frac{1}{N}\frac{\partial
N}{\partial\triangle}\Bigg),\label{k2eq}\\
K_3&=&\frac{K_1}{8(\triangle-1)^2}\Bigg(-2\frac{K_2}{K_1}
+\ln^2[f]+\frac{2}{N}\frac{\partial
N}{\partial\triangle}\ln[f]+\frac{1}{N}
\frac{\partial^2N}{\partial\triangle^2}\Bigg).\label{k3eq}
\end{eqnarray}
We are now in a position to calculate two-point correlation
functions from the equations \eqref{eq1}-\eqref{eq3}.  For this
purpose,  we  consider an on-shell bilinear action $ S_{{\rm eff}}$,
obtained from surface integral on the boundary $s$ after some
integration by parts,
\begin{eqnarray}\label{seff}
&& S_{{\rm eff}}[A_{1(b)},A_{2(b)},A_{3(b)}]\nonumber\\
&&\hspace*{-0.5em}=-\lim_{r_s\to\infty}
\frac{1}{2}\int_{s}du_+du_-\sqrt{-\gamma}\Big[A_1(\hat{n}\cdot\nabla)
A_3 +A_2(\hat{n}\cdot\nabla) A_2+A_3(\hat{n}\cdot\nabla) A_1\Big],
\end{eqnarray}
where $s$ is a regulated surface at $r=r_s$, a normal derivative
$\hat{n}\cdot\nabla=r\partial_r$, and $\gamma_{\mu\nu}$ is the
induced metric on the boundary.  We notice that on the boundary,
$A_{i}$ behave as $A_{i}|_{r\to\infty}\sim
r^{-2+\triangle}A_{i(b)}$. In this case, substituting \eqref{a3}
together with \eqref{k1eq}-\eqref{k3eq} into the effective action
\eqref{seff} leads to
\begin{eqnarray}
 S_{{\rm eff}}&=&\frac{\triangle N}{2}\int
du_+du_-du_+^{\prime}du_-^{\prime}\left(\frac{\pi T_L}{\sinh[\pi
T_L\delta u_+]}\right)^{\triangle}\left(\frac{\pi T_R}{\sinh[\pi
T_R\delta u_-]}\right)^{\triangle}\times\nonumber\\
&&\Bigg\{A_{1(b)}(u)A_{3(b)}(u^{\prime})+A_{2(b)}(u)A_{2(b)}(u')
+A_{3(b)}(u)A_{1(b)}(u')\nonumber\\
&&+\frac{1}{2(\triangle-1)}
\left(\triangle^{-1}+\frac{1}{N}\frac{\partial
N}{\partial\triangle}+\ln[\epsilon]+\ln\left[\frac{\pi^2
T_RT_L}{\sinh[\pi T_L\delta u_+]\sinh[\pi T_R\delta
u_-]}\right]\right)\times\nonumber\\
&&\Big(A_{1(b)}(u)A_{2(b)}(u')+A_{2(b)}(u)A_{1(b)}(u')\Big)
\nonumber\\
&&+\frac{1}{8(\triangle-1)^2}
\left(\frac{1}{(1-\triangle)N}\frac{\partial
N}{\partial\triangle}+\frac{1}{N}\frac{\partial^2N}{\partial\triangle^2}
+\frac{2}{\triangle}\ln\left[\frac{\pi^2 T_RT_L}{\sinh[\pi T_L\delta
u_+]\sinh[\pi T_R\delta
u_-]}\right]\epsilon\nonumber\right.\\
&&\left.+\ln^2\left[\frac{\pi^2 T_RT_L}{\sinh[\pi T_L\delta
u_+]\sinh[\pi T_R\delta u_-]}\right]\epsilon^2\right)
A_{1(b)}(u)A_{1(b)}(u')\Bigg\}, \label{bieff}
\end{eqnarray}
where $\epsilon=1/r_s\to0$ as $r_s$ goes infinity.

Following the AdS/LCFT correspondence, we couple the boundary values of the fields to the dual
operators as
\begin{equation}
\int d^2x \Bigg[\sum_{i=1}^3 A_{4-i(b)}{\cal O}_i\Bigg]
\end{equation}
for the Jordan-type coupling.  One can derive the two-point
functions for the dual conformal operators ${\cal O}_{i}$ as
follows:
\begin{eqnarray}
\Big<{\cal O}_{1}(u_+,u_-){\cal O}_{1}(0)\Big>
&=&\frac{\delta^2S_{eff}}{\delta A_{3(b)}(u_+,u_-)\delta
A_{3(b)}(0)}~=~0,\label{o11}\\
\Big<{\cal O}_{1}(u_+,u_-){\cal O}_{2}(0)\Big>
&=&\frac{\delta^2S_{eff}}{\delta A_{3(b)}(u_+,u_-)\delta
A_{2(b)}(0)}~=~0,\label{o12}\\
\nonumber\\
\xi\Big<{\cal O}_{1}(u_+,u_-){\cal O}_{3}(0)\Big>
&=&\frac{\delta^2S_{eff}}{\delta A_{3(b)}(u_+,u_-)\delta
A_{1(b)}(0)}\nonumber\\
\hspace*{9.2em} &=& \triangle N\left(\frac{\pi T_L}{\sinh[\pi
T_Lu_+]}\right)^{\triangle} \left(\frac{\pi T_R}{\sinh[\pi T_R
u_-]}\right)^{\triangle},
\label{o13}\\
\nonumber\\
\zeta\Big<{\cal O}_{2}(u_+,u_-){\cal O}_{2}(0)\Big>
&=&\frac{\delta^2S_{eff}}{\delta A_{2(b)}(u_+,u_-)\delta
A_{2(b)}(0)}\nonumber\\
&=& \frac{\triangle N}{2}\left(\frac{\pi T_L}{\sinh[\pi
T_Lu_+]}\right)^{\triangle} \left(\frac{\pi T_R}{\sinh[\pi T_R
u_-]}\right)^{\triangle},\label{o22}\\
\nonumber
\end{eqnarray}
\begin{eqnarray}
 \eta\Big<{\cal O}_{2}(u_+,u_-){\cal O}_{3}(0)\Big>
&=&\frac{\delta^2S_{eff}}{\delta A_{2(b)}(u_+,u_-)\delta
A_{1(b)}(0)}\nonumber\\
&=& \frac{\triangle N}{2(\triangle-1)}\left(\frac{\pi T_L}{\sinh[\pi
T_Lu_+]}\right)^{\triangle} \left(\frac{\pi T_R}{\sinh[\pi T_R
u_-]}\right)^{\triangle}\times\nonumber\\
&&\left(\frac{1}{\triangle}+\frac{1}{N}\frac{\partial
N}{\partial\triangle}+\ln\left[\frac{\pi^2 T_RT_L}{\sinh[\pi T_L
u_+]\sinh[\pi T_R u_-]}\right]\epsilon\right),\label{o23}\\
\nonumber\\
\chi \Big<{\cal O}_{3}(u_+,u_-){\cal O}_{3}(0)\Big>
&=&\frac{\delta^2S_{eff}}{\delta A_{1(b)}(u_+,u_-)\delta
A_{1(b)}(0)}\nonumber\\
&=& \frac{\triangle N}{16(\triangle-1)^2}\left(\frac{\pi
T_L}{\sinh[\pi T_Lu_+]}\right)^{\triangle} \left(\frac{\pi
T_R}{\sinh[\pi T_R
u_-]}\right)^{\triangle}\times\nonumber\\
&&\hspace*{-3em}\left\{\frac{\triangle}{(1-\triangle)N}\frac{\partial
N}{\partial\triangle}+\frac{\triangle}{N}\frac{\partial^2
N}{\partial\triangle^2}+\frac{2}{\triangle}\ln\left[\frac{\pi^2
T_RT_L}{\sinh[\pi T_L u_+]\sinh[\pi T_R
u_-]}\right]\epsilon\right.\nonumber\\
&&\left.+\ln^2\left[\frac{\pi^2 T_RT_L}{\sinh[\pi T_L u_+]\sinh[\pi
T_R u_-]}\right]\epsilon^2\right\}\label{o33}
\end{eqnarray}
where arbitrary parameters $\xi,~\zeta,~\eta,$ and $\chi$ will be
determined when reducing to the AdS$_3$ background. This is a rank-3
finite temperature LCFT, our main result.

According to the AdS/CFT dictionary, there are two possibilities of
choosing  coupling to boundary operators: the first one is symmetric
and the second one is Jordan (non-symmetric). The first provides a
standard coupling of $\int d^2x[\sum_{i=1}^3 A_{i(b)}{\cal O}_i]$,
while the second gives us the coupling of $\int d^2x[A_{3(b)}{\cal
O}_1+A_{1(b)}{\cal O}_3+A_{2(b)}{\cal O}_2]$. Here, we choose the
Jordan  case which makes the coupling between boundary field
$A_{i(b)}$ and operator ${\cal O}_i$ less transparent. This implies
that in order to obtain the  two-point correlator $<{\cal O}_3{\cal
O}_3>$, one has to vary the effective action (\ref{bieff}) with
respect to $A_{1(b)}$ twice, but $<{\cal O}_1{\cal O}_1>$  is
obtained  when varying with respect to $A_{3(b)}$ twice.

\section{Rank-3 LCFT on AdS$_3$}

 One can check
that in the low temperature limit of $T_{R/L}\to0$, two-point
functions\footnote{We mention that log-term in Eq. (\ref{k3eq}) can
be eliminated by adding $K_2$ to $K_3$~\cite{Bergshoeff:2012sc}. In
this case, the correlation function (\ref{o33}) is changed
slightly.} for $\Big\{{\cal O}_1,~{\cal O}_2,~{\cal O}_3\Big\}$
reduce to those for
 $\Big\{{\cal O}^{{\rm KG}},~{\cal O}^{{\rm
log}},~{\cal O}^{{\rm log}^2}\Big\}$  defined in the Euclidean Poincar\'{e} patch of AdS$_{3}$
\begin{equation}
ds^2_{\rm EAdS}=\frac{1}{z^2}\Big(dz^2+dx_1^2+dx_2^2\Big).
\end{equation}
These are~\cite{Bergshoeff:2012sc}
\begin{eqnarray}
\label{ac1} &&\Big<{\cal O}^{{\rm KG}}(x){\cal O}^{{\rm
KG}}(0)\Big>=\Big<{\cal
O}^{{\rm KG}}(x){\cal O}^{{\rm log}}(0)\Big>=0,\\
\label{ac2}&&\Big<{\cal O}^{{\rm KG}}(x){\cal O}^{{\rm
log}^2}(0)\Big>=\Big<{\cal O}^{{\rm log}}(x){\cal O}^{{\rm
log}}(0)\Big>=
\frac{\Big(2(\triangle-1)\Big)^3}{|x|^{2\triangle}},\\
\label{ac3}&&\Big<{\cal O}^{{\rm log}}(x){\cal O}^{{\rm
log^2}}(0)\Big>=\frac{\Big(2(\triangle-1)\Big)^3}{|x|^{2\triangle}}\Big(-2\ln|x|
+\Lambda_1\Big),\\
\label{ac4}&&\Big<{\cal O}^{{\rm log}^2}(x){\cal O}^{{\rm
log}^2}(0)\Big>=\frac{\Big(2(\triangle-1)\Big)^3}{|x|^{2\triangle}}
\Big(2\ln^2|x| +\Lambda_1\ln|x|+\Lambda_2\Big),
\end{eqnarray}
where $|x|=\sqrt{x_1^2+x_2^2}$ and constants $\Lambda_{1,2}$ are
related to the arbitrary shift parameters $\lambda_{1,2}$, given by
the shift transformation: $\phi_i\to
\phi_i+\sum_{k=1}^{i-1}\lambda_k \phi_{i-k}$ ($i=2,3$).\\ From this
reduction, $N,~\xi,~\zeta,~\eta,~\chi$, $\Lambda_1,$ and $\Lambda_2$
are determined to be\footnote{Note that considering the shift
transformation of $A_i\to A_i+\sum_{k=1}^{i-1}\tilde\lambda_k
A_{i-k}$ ($i=2,3$) with arbitrary parameters $\tilde\lambda$, two
parameters $\Lambda_1$ and $\Lambda_2$  remain  undetermined as
functions of ($\tilde\lambda_1,\tilde\lambda_2$).}
\begin{eqnarray}
\xi=\frac{\triangle
N}{\Big(2(\triangle-1)\Big)^3},~~~\zeta=\frac{\triangle
N}{2\Big(2(\triangle-1)\Big)^3},~~~\eta=\frac{\triangle
N}{\Big(2(\triangle-1)\Big)^4},~~~\chi=\frac{\triangle
N}{2\Big(2(\triangle-1)\Big)^5}
\end{eqnarray}
\begin{eqnarray}
\Lambda_1=\frac{1}{\triangle}+\frac{1}{N}\frac{\partial
N}{\partial\triangle},
~~~\Lambda_2=\frac{\triangle}{2N}\frac{\partial^2
N}{\partial\triangle^2}-\frac{\triangle}{N^2}\left(\frac{\partial
N}{\partial\triangle}\right)^2,
\end{eqnarray}
where $N=c\triangle^{-3}$ with an integration constant $c$. The
connection between $u_\pm$ and $x_{1/2}$ is given by
\begin{equation}
u_+=x_1+ix_2,~~u_-=x_1-ix_2. \end{equation}

Actually, Eqs.(\ref{ac1})-(\ref{ac4}) correspond to the correlation
functions for the un-truncated rank-3 LCFT model. For odd rank, the
theory has a unitary subspace which could be obtained by truncating
half of the logarithmic modes.  An easy way to see how this works
for a  rank-3 case  is first to rewrite the two-point correlation
functions schematically as \begin{equation}
 <{\cal O}^i{\cal O}^j>
\sim \left(
  \begin{array}{ccc}
    0 & 0 & {\rm CFT} \\
    0 & {\rm CFT} & {\rm L} \\
   {\rm  CFT} & {\rm L} & {\rm L}^2 \\
  \end{array}
\right), \label{3by3cft}
 \end{equation}
 where $i,j$ =KG, log, log$^2$, CFT denotes the CFT two-point
 function (\ref{ac2}), L represents log two-point function (\ref{ac3}), and L$^2$ denotes
 log-square two-point function (\ref{ac4}).
 If one truncates the theory to be unitary, one throws away all
 modes which generates the third column and row of the above matrix.
 The only non-zero correlation function is given by a reduced
 matrix as
 \begin{equation}
  \label{tmatrix}
 <{\cal O}^i{\cal O}^j> \sim \left(
  \begin{array}{cc}
    0& 0 \\
    0 & {\rm CFT} \\
  \end{array}
\right),
 \end{equation}
 which implies that the remaining   sector involves a non-trivial
 two-point correlator
 \begin{equation}
 <{\cal O}^{\rm log}(x){\cal O}^{\rm log}(0)>=\frac{\Big[2(\triangle-1)\Big]^3}{|x|^{2\triangle}}. \end{equation}
 This is unitary and thus, the non-unitary issue could be  resolved  by truncating  a rank-3 LCFT.

\section{Retarded Green's functions of rank-3 LCFT}
The AdS/LCFT correspondence implies that quasinormal frequencies
$\omega=\omega_r-i\omega_i$
 determining the relaxation times of
perturbations $A_i$ around the BTZ black hole  are agreed with the
location of poles of the retarded Green's function of the
corresponding perturbations ${\cal O}_{i}$ in the dual
CFT~\cite{Birmingham:2001pj}. The quasinormal modes are obtained by
imposing the boundary conditions: ingoing waves near the horizon and
the Dirichlet condition at infinity  because in asymptotically AdS
spaces, the spacelike infinity acts like a reflecting boundary.  The
regime of near equilibrium real-time evolution  can be captured via
the linear response theory.   A major role is being played by the
pole of these holographic response functions (quasinormal
frequencies). Actually, the retarded Green's functions are the
central objects in linear response theory which are defined by
\begin{equation} \label{retgreen}
{\cal D}^{\rm ret}_{jk}(t,\phi;0,0)=i\Theta(t-0)\bar{\cal
D}_{jk}(u_+,u_-),~~j,k=1,2,3,
\end{equation}
where  the  commutator evaluated in the equilibrium canonical
ensemble are given by
\begin{eqnarray}
\bar{\cal D}_{jk}(u_+,u_-)=\Big<{\cal O}_{j}(u_+ -i\epsilon,-u_-
-i\epsilon){\cal O}_{k}(0)\Big>-\Big<{\cal O}_{j}(u_+
+i\epsilon,-u_- +i\epsilon){\cal O}_{k}(0)\Big>.
\end{eqnarray}
Making  the Fourier transform of  $\bar{\cal D}_{jk}(u_+,u_-)$, the
commutator $\bar{\cal D}_{jk}(p_+,p_-)$ takes the form
\begin{eqnarray}\label{Dinp}
\bar{\cal D}_{jk}(p_+,p_-)=\int du_+ du_- e^{i(p_+ u_+-p_-
u_-)}\bar{\cal D}_{jk}(u_+,u_-)
\end{eqnarray}
in the momentum space. Here $p_\pm=(\omega\mp k)/2$.
 Using (\ref{Dinp}) and
\eqref{o11} [\eqref{o12}],  one  finds  that
\begin{eqnarray}\label{d11}
\bar{\cal D}_{11}(p_+,p_-)=\bar{\cal D}_{12}(p_+,p_-)=0.
\end{eqnarray}
Plugging \eqref{o13} [\eqref{o22})] into (\ref{Dinp}), we
 obtain the commutator $\bar{\cal D}_{13}(p_+,p_-)$
[$\bar{\cal D}_{22}(p_+,p_-)$],
\begin{eqnarray}
&&\bar{\cal D}_{13}(p_+,p_-)~=~\bar{\cal
D}_{22}(p_+,p_-)\nonumber\\
&=&\Big(2(\triangle-1)\Big)^3\int du_+ du_- e^{ip_+ u_+}e^{-ip_-
u_-}\Bigg\{\left(\frac{\pi T_L}{\sinh[\pi
T_L(u_+-i\epsilon)]}\right)^{\triangle}\times\nonumber\\
&&\hspace*{-2em}\left(\frac{\pi T_R}{\sinh[\pi T_R
(u_--i\epsilon)]}\right)^{\triangle}-\left(\frac{\pi T_L}{\sinh[\pi
T_L(u_++i\epsilon)]}\right)^{\triangle} \left(\frac{\pi
T_R}{\sinh[\pi T_R (u_-+i\epsilon)]}\right)^{\triangle}\Bigg\}
\nonumber\\
&&\nonumber\\
&=&\Big(2(\triangle-1)\Big)^3\frac{(2\pi T_L)^{\triangle-1}(2\pi
T_R)^{\triangle-1}}{\Gamma(\triangle)^2} \sinh\left[\frac{p_+}{2T_L}
+\frac{p_-}{2T_R}\right]\times\nonumber\\
&&\hspace*{8em}\Big|
\Gamma\left(\frac{\triangle}{2}+i\frac{p_+}{2\pi T_L}\right)\Big|^2
\Big| \Gamma\left(\frac{\triangle}{2}+i\frac{p_-}{2\pi
T_R}\right)\Big|^2\label{d13}
\end{eqnarray}
where $\Gamma$ is the gamma function. In deriving this, we have used the formula
\begin{equation}
\int dx e^{-i\omega x}(-1)^{\triangle}\left(\frac{\pi T}{\sinh[\pi
T(x\pm\epsilon)]}\right)^{2\triangle}~=~~\frac{(2\pi
T)^{2\triangle-1}}{\Gamma(2\triangle)} e^{\pm \omega/2T}
\Big|\Gamma\left(\triangle+i\frac{\omega}{2\pi T}\right)\Big|^2.
\end{equation}

At this stage, we wish to  point out that it is not easy to compute
the commutators $\bar{\cal D}_{23}(p_+,p_-)$ and  $\bar{\cal
D}_{33}(p_+,p_-)$ directly  because they have the logarithmic
singularities as are shown in  (\ref{o23}) and (\ref{o33}). It is
suggested, however, that $\bar{\cal D}_{23}$ can be deduced from the
relation \cite{Myung:1999nd}
\begin{eqnarray}\label{o23o13}
\Big<{\cal O}_{2}(u_+,u_-){\cal
O}_{3}(0)\Big>=\frac{\partial}{\partial\triangle}\Big<{\cal
O}_{1}(u_+,u_-){\cal O}_{3}(0)\Big>.
\end{eqnarray}
We compute  $\bar{\cal D}_{23}(p_+,p_-)$
by using the  relation \eqref{o23o13}, which yields
\begin{eqnarray}\label{d23d13}
\hspace*{-2em}\bar{\cal
D}_{23}(p_+,p_-)&=&\Bigg\{\frac{3}{\triangle-1}+\ln[2\pi
T_L]+\ln[2\pi
T_R]-2\psi(\triangle)+\frac{1}{2}\psi\left(\frac{\triangle}{2}
+i\frac{p_+}{2\pi T_L}\right)\nonumber\\
&&\hspace*{-6.7em}+\frac{1}{2}\psi\left(\frac{\triangle}{2}
-i\frac{p_+}{2\pi T_L}\right)
+\frac{1}{2}\psi\left(\frac{\triangle}{2} +i\frac{p_-}{2\pi
T_R}\right) +\frac{1}{2}\psi\left(\frac{\triangle}{2}
-i\frac{p_-}{2\pi T_R}\right)\Bigg\}\bar{\cal D}_{13}(p_+,p_-)  \\
&\equiv& \Psi(p_+,p_-)\bar{\cal D}_{13}(p_+,p_-),\nonumber
\end{eqnarray}
where $\psi(A)=\partial\ln[\Gamma(A)]/\partial A$ is the digamma
function. Importantly, we may calculate  $\bar{\cal D}_{33}$
when using the relation
\begin{eqnarray}\label{o33o23}
\Big<{\cal O}_{3}(u_+,u_-){\cal
O}_{3}(0)\Big>=\frac{1}{2}\frac{\partial^2}{\partial\triangle^2}\Big<{\cal
O}_{1}(u_+,u_-){\cal O}_{3}(0)\Big>.
\end{eqnarray}
Accordingly,  $\bar{\cal D}_{33}$ takes the form in the momentum space
\begin{eqnarray}\label{d33d13}
\bar{\cal
D}_{33}(p_+,p_-)&=&\Bigg\{-\frac{3}{2(\triangle-1)^2}-\psi'(\triangle)
+\frac{1}{8}\psi'\left(\frac{\triangle}{2} +i\frac{p_+}{2\pi
T_L}\right)+\frac{1}{8}\psi'\left(\frac{\triangle}{2}
-i\frac{p_+}{2\pi T_L}\right)\nonumber\\
&&\hspace*{-3.7em} +\frac{1}{8}\psi'\left(\frac{\triangle}{2}
+i\frac{p_-}{2\pi T_R}\right)
+\frac{1}{8}\psi'\left(\frac{\triangle}{2} -i\frac{p_-}{2\pi
T_R}\right)+\frac{1}{2}\Psi(p_+,p_-)^2\Bigg\}\bar{\cal
D}_{13}(p_+,p_-),
\end{eqnarray}
where  the prime (${}^{\prime}$) denotes the differentiation with
respect to $\triangle$.

It is worth noting  that for the non-rotating BTZ black hole
($T_R=T_L=T_H=1/2\pi \ell,~r_+=\ell)$,  the pole structure of the
commutators $\bar{\cal D}_{13}(p_+)=\bar{\cal
D}_{22}(p_+),~\bar{\cal D}_{23}(p_+)$, and $\bar{\cal D}_{33}(p_+)$
 are given by (after setting $\triangle=2h_L$)
\begin{eqnarray}
\label{bac1}\bar{\cal
D}_{13}(p_+)&\propto&\Gamma\left(h_L+i\frac{p_+}{2\pi
T_L}\right)\Gamma\left(h_L-i\frac{p_+}{2\pi T_L}\right),\\
\nonumber \bar{\cal
D}_{23}(p_+)&\propto&\Gamma'\left(h_L+i\frac{p_+}{2\pi
T_L}\right)\Gamma\left(h_L-i\frac{p_+}{2\pi T_L}\right) \\
\label{bac2} &&+\Gamma\left(h_L+i\frac{p_+}{2\pi
T_L}\right)\Gamma'\left(h_L-i\frac{p_+}{2\pi T_L}\right),\\
\nonumber \bar{\cal
D}_{33}(p_+)&\propto&\Gamma''\left(h_L+i\frac{p_+}{2\pi
T_L}\right)\Gamma\left(h_L-i\frac{p_+}{2\pi T_L}\right) \nonumber \\
\nonumber &&+2\Gamma'\left(h_L+i\frac{p_+}{2\pi
T_L}\right)\Gamma'\left(h_L-i\frac{p_+}{2\pi T_L}\right)\\
\label{bac3} &&+\Gamma\left(h_L+i\frac{p_+}{2\pi
T_L}\right)\Gamma''\left(h_L-i\frac{p_+}{2\pi T_L}\right),
\end{eqnarray}
which are consistent with those obtained  in the literatures
\cite{Birmingham:2001pj,Sachs:2008yi,KMP}. For a tricritical GMG, we
read off  the log-square quasinormal frequencies of graviton with
$h_L=0$
\begin{equation}
\omega_t = k- i4\pi T_L (n+h_L). \label{10}
\end{equation}
 from a triple pole of the retarded Green's
function in the momentum space  $\bar{\cal D}_{33}(p_+)$ .

 We represent  the
whole pole structures as a three-by-three matrix
 \begin{equation} \label{ftm}
 \bar{\cal D}_{ij}
\sim \left(
  \begin{array}{ccc}
    0 & 0 & \bar{\cal D}_{13} \\
    0 & \bar{\cal D}_{22} & \bar{\cal D}_{23} \\
   \bar{\cal D}_{13} & \bar{\cal D}_{23} & \bar{\cal D}_{33} \\
  \end{array}
\right),
 \end{equation}
 where $\bar{\cal D}_{13}[=\bar{\cal D}_{22}]$ represent a simple pole, $\bar{\cal D}_{23}$ denotes a double pole,
 and $\bar{\cal D}_{33}$ represents a triple pole in the lower-half
 plane. These pole structures are the same  as the retarded Green's
 function (\ref{retgreen}).

What happens for the quasinormal modes when truncating the
log-square quasinormal modes on the BTZ black hole~\cite{KMP}? At
this stage, we may  answer to this question because we did construct
all retarded Green's functions. Applying  the previous truncation
process done for the LCFT to (\ref{ftm}) leads to
\begin{equation}
  \label{tmatrix}
 \bar{\cal D}_{ij} \sim \left(
  \begin{array}{cc}
    0& 0 \\
    0 & \bar{\cal D}_{22} \\
  \end{array}
\right),
 \end{equation}
which provides a simple pole  existing in the finite temperature
CFT. From (\ref{bac1}), one reads off  quasinormal frequencies as
\begin{equation}
\omega_s = k- i4\pi T_L (n+h_L). \label{10}
\end{equation}

\section{Absorption cross sections}
In this section, we wish to calculate the $s$-wave absorption cross
section (or greybody factor) by  using the retarded Green's
functions
 in the momentum spaces.  Actually, it is not an
easy task to derive all absorption cross sections by solving
equations of motion directly  because two scalars $A_2$ and $A_3$
satisfy the higher-order differential equations (\ref{aa2}) and
(\ref{aa3}) on the BTZ black hole.  However, when using the boundary
retarded Green's  functions, one  computes the absorption cross
sections easily.

It is well known that the absorption cross section
\cite{Gubser:1997cm,Teo:1998dw,MullerKirsten:1998mt} can be written
in terms of frequency ($\omega$) and temperature ($T_{R/L},~T_H$):
\begin{eqnarray}\label{abs0}
\sigma_{\rm abs}^{ij} &=&\frac{{\cal C}_0}{\omega}\bar{\cal
D}_{ij}(\omega),
\end{eqnarray}
where ${\cal C}_0$ is a normalization constant. Here $\bar{\cal
D}_{ij}(\omega)$ is obtained  by substituting $p_+=p_-=\omega/2$ for
$s$-wave ($k=0$) into  (\ref{d11}), (\ref{d13}), (\ref{d23d13}), and
(\ref{d33d13}). From the  definition (\ref{abs0}), one can find
$\sigma_{\rm abs}^{13}$ as
\begin{eqnarray}
&&\sigma_{\rm abs}^{13}= \sigma_{\rm abs}^{22}= \frac{{\cal
C}_0\Big(2(\triangle-1)\Big)^3 (2 \pi T_L\ell)^{\Delta -1} (2 \pi
T_R \ell)^{\Delta -1}
    \sinh({\omega \over 2 T_H})}{
\omega \Gamma^2(\Delta)}
\nonumber \\
&&\hspace*{7em}\times \left \vert \Gamma \left ({\Delta \over 2} + i
{\omega \over 4 \pi T_L} \right )
          \Gamma\left ( {\Delta \over 2} + i {\omega \over 4 \pi T_R}
           \right )
\right \vert^2 \label{abs13}
\end{eqnarray}
where  the AdS$_3$ curvature radius $\ell=1$ is restored for
convenience. For
 $\Delta=2~(m^2=0)$, $\sigma_{\rm abs}^{13}$  is  the same with the
absorption cross section for a massless minimally coupled scalar
which satisfies $\nabla^2_B A_1 = 0$ \cite{Birmingham:1997rj,Lee:1998pj}
\begin{equation}
\sigma_{\rm abs}^{13} =\sigma_{\rm abs}^{22}= \pi^2 \omega \ell^2
{e^{\omega/T_H} -1 \over \left ( e^{\omega/2 T_L} - 1 \right ) \left
( e^{\omega/2 T_R} - 1 \right )} \label{abs-minimal}
\end{equation}
with ${\cal C}_0=1/4$. One can easily check that in the low-energy
limit of $\omega \ll T_{R/L}$, the absorption cross section
$\sigma_{\rm abs}^{13}$ reduces to the area-law of $\sigma_{\rm
abs}^{13} \vert_{\omega \ll T_{R/L}} =$ $ 2 \pi r_+ = $ ${\cal
A}_H$, while in the low-temperature limit of $\omega \gg T_{R/L}$,
it becomes $\sigma_{\rm abs}^{13} \vert_{\omega \gg T_{R/L}} =$
$\pi^2 \omega \ell^2$.

On the other hand,  from the relation of (\ref{d23d13}),  we find the absorption cross section
\begin{eqnarray}\label{a23a13}
\sigma_{\rm abs}^{23} &\simeq& \Bigg [ {3 \over \Delta-1 } + \ln(2
\pi T_L\ell) + \ln(2 \pi T_R\ell) - 2 \psi(\Delta) +{1 \over 2}
\left \{
\psi({\Delta \over 2}+i{\omega \over 4 \pi T_L})\right.\nonumber\\
&&\left. + \psi({\Delta \over 2}-i{\omega \over 4 \pi T_L}) +
\psi({\Delta \over 2}+i{\omega \over 4 \pi T_R}) + \psi({\Delta
\over 2}-i{\omega \over 4 \pi T_R}) \right \} \Bigg ]\sigma_{\rm
abs}^{13}.
\end{eqnarray}
Taking  the low-temperature limit of $\omega \gg T_{R/L}=1/2\pi
\ell$ and $\Delta=2$,  we get the absorption cross section which is
logarithmically corrected  as~\cite{Myung:1999nd}
\begin{equation}
\sigma_{\rm abs}^{23}\vert_{\omega \gg T_{R/L}} = { \pi^2 \omega
\ell^2 } \Big(
           1 + 2\ln[\omega \ell]+c_1\Big),
\label{abs-semifinal}
\end{equation}
where $c_1$ is given by
\begin{eqnarray}
c_1=2\gamma-2\ln2
\end{eqnarray}
 with the Euler's constant $\gamma=0.5772$. The absorption cross
 section \eqref{abs-semifinal} is positive definite for the low-temperature limit because $1+2\gamma-2\ln2\simeq 0.768$. In
deriving  Eq.\eqref{abs-semifinal}, we have used an asymptotic form
of the digamma function
\begin{eqnarray}
{\rm Re} [\psi(1+ix)] = {\rm Re} [\psi(1-ix)] &\simeq& \ln x + {1
\over 12 x^2} + O(x^{-4}).
\end{eqnarray}
We note that Eq.(\ref{abs-semifinal}) shows the absorption cross
section for the logarithmic operator.

Finally,  Eq.(\ref{d33d13}) provides the absorption cross section for the log-square operator
\begin{eqnarray}
\sigma_{\rm abs}^{33} &\simeq&
\Bigg\{-\frac{3}{2(\triangle-1)^2}-\psi'(\triangle)
+\frac{1}{8}\psi'\left(\frac{\triangle}{2} +i\frac{\omega}{4\pi
T_L}\right)+\frac{1}{8}\psi'\left(\frac{\triangle}{2}
-i\frac{\omega}{4\pi T_L}\right)\nonumber\\
&&\hspace*{-3.7em} +\frac{1}{8}\psi'\left(\frac{\triangle}{2}
+i\frac{\omega}{4\pi T_R}\right)
+\frac{1}{8}\psi'\left(\frac{\triangle}{2} -i\frac{\omega}{4\pi
T_R}\right)+\frac{1}{2}\Psi(\omega)^2\Bigg\}\sigma_{\rm abs}^{13},
\nonumber
\end{eqnarray}
where the last term is obtained  through the relation of
$\sigma_{\rm abs}^{23}\equiv \Psi(\omega)\sigma_{\rm abs}^{13}$
defined in Eq.(\ref{a23a13}). We consider  the low-temperature limit
of $\omega \gg T_{R/L}$ and $\Delta=2$. In this case,  $\psi'$ takes
the asymptotic form
\begin{eqnarray}
{\rm Re} [\psi'(1+ix)] = {\rm Re} [\psi'(1-ix)] &\simeq& {1 \over 2
x^2} + O(x^{-5}).
\end{eqnarray}
Then, $\sigma_{\rm abs}^{33}$ leads to the form
\begin{equation}
\sigma_{\rm abs}^{33}\vert_{\omega \gg T_{R/L}} = { \pi^2 \omega
\ell^2 } \Big(
           1 +2\ln^2[\omega \ell] +c_2\ln[\omega \ell]+ c_3 \Big),
\label{abs-final}
\end{equation}
where $c_2$ and $c_3$ are given by
\begin{eqnarray}
c_2&=&2(1+2\gamma-2\ln2),\nonumber\\
c_3&=&-\frac{1}{6}\pi^2+2\gamma^2+2\gamma-2(1+2\gamma)\ln2
+2(\ln2)^2-1.
\end{eqnarray}
Note that the absorption cross section \eqref{abs-final} is positive
definite for $c_2\ln[\omega\ell]+c_3+1>0$ because $c_2\simeq1.536>0$
and  $c_3+1\simeq-1.849$.

We represent  the whole cross sections as a three-by-three matrix
 \begin{equation} \label{abstm}
 \sigma^{ij}_{\rm abs}
\sim \left(
  \begin{array}{ccc}
    0 & 0 & \sigma^{13}_{\rm abs} \\
    0 &\sigma^{22}_{\rm abs} &\sigma^{23}_{\rm abs} \\
   \sigma^{13}_{\rm abs} & \sigma^{23}_{\rm abs} & \sigma^{33}_{\rm abs} \\
  \end{array}
\right),
 \end{equation}
 where $\sigma^{13}_{\rm abs}[=\sigma^{22}_{\rm abs}]$ represents the absorption cross section
 when  the Klein-Gordon mode is scattered by the BTZ black hole,
 $\sigma^{23}_{\rm abs}$  denotes the absorption cross section
 when  the log-mode is scattered by the BTZ black hole,
 and $\sigma^{33}_{\rm abs}$ represents the absorption cross section
 when  the log-square mode is scattered by the BTZ black hole.
Applying  the previous truncation process done for the LCFT to
(\ref{abstm}) leads to
\begin{equation}
  \label{tmatrix}
 \sigma^{ij}_{\rm abs} \sim \left(
  \begin{array}{cc}
    0& 0 \\
    0 & \sigma^{22}_{\rm abs}\\
  \end{array}
\right),
 \end{equation}
which provides the absorption cross section for the Klein-Gordon
mode in the low-temperature limit.

\section{Summary and conclusion}
We have constructed a rank-3 finite temperature logarithmic
conformal field theory (LCFT)  starting from a higher-derivative
scalar field model in the BTZ black hole background.

 On the gravity
side, this scalar field model is composed of two auxiliary scalar
fields $A_1$ and $A_2$ and one relevant field $A_3$. The tricritical
point where the masses of the three scalars degenerate was
introduced to avoid the ghost problem. It is known that the ghost
issue was captured even in the Minkowski space for the non-critical
case  of a six-derivative scalar theory [see appendix C in the
Ref.\cite{Barth:1983hb}]. Since the scalar $A_2$ satisfies the
fourth-order differential equation (\ref{aa2}) and $A_3$ satisfies
the sixth-order differential equation (\ref{aa3}), one could expect
that two higher-order logarithmic modes appear and these correspond
to two logarithmic partners of the Klein-Gordon scalar mode.
According to the AdS/CFT logic, there are two possibilities of
coupling to boundary operators: the first one is symmetric and the
second one is Jordan. The first provides a standard coupling of
$\int d^2x[\sum_{i=1}^3 A_{i(b)}{\cal O}_i]$, while the second gives
us the coupling of $\int d^2x[A_{3(b)}{\cal O}_1+A_{1(b)}{\cal
O}_3+A_{2(b)}{\cal O}_2]$.  In this work, we have chosen the Jordan
which makes the coupling between boundary field $A_{i(b)}$ and
operator ${\cal O}_i$ less transparent. This implies that in order
to obtain two-point correlator $<{\cal O}_3{\cal O}_3>$, one has to
vary the effective action (\ref{bieff}) with respect to $A_{1(b)}$
twice, but $<{\cal O}_1{\cal O}_1>$  with respect to $A_{3(b)}$
twice. This could be also  observed from (\ref{eeq3}) obtained when
varying (\ref{maina}) with respect to $\Phi_1$ but not $\Phi_3$.

On the CFT side,  all correlators describe  a rank-three finite
temperature LCFT. Its zero temperature limit reduces to  a rank-3
LCFT  which is dual of a higher-derivative scalar field model in the
AdS$_3$ background~\cite{Bergshoeff:2012sc}. Here, a truncation
allowed the theory to have a unitary subspace by throwing away all
modes which generate the third column and row of (\ref{3by3cft}).

 We have computed the retarded real-time Green's
functions and retarded Green's functions in the momentum space to
know the locations of poles. Then, the AdS/LCFT correspondence
implies that quasinormal frequencies determining the relaxation
times of perturbations around the black hole  are agreed with the
location of poles of the retarded Green's function of the
corresponding perturbations in the dual CFT. For a tricritical GMG,
we read off the log-square quasinormal frequencies of graviton from
the poles of the retarded Green's function in the momentum space.
This confirms the previously approximate computation~\cite{KMP}.
After implementing the truncation process, we have found quasinormal
frequencies from a unitary conformal field theory.

Finally, we have obtained  the absorption cross sections by using
the retarded Green's functions on the boundary. Actually, it is not
an easy task to derive all absorption cross sections because two
scalars $A_2$ and $A_3$ satisfy the higher-order differential
equations (\ref{aa2}) and (\ref{aa3}). However, the boundary
retarded Green's functions allow us to compute the absorption cross
sections easily. In the low-temperature limit of $\omega \gg
T_{R/L}$, one has $\sigma^{13}_{\rm abs}=\sigma^{22}_{\rm abs}=\pi^2
\omega \ell^2$, $\sigma^{23}_{\rm abs}=\pi^2 \omega
\ell^2(1+2\ln[\omega \ell]+c_1)$, and $\sigma^{33}_{\rm abs}=\pi^2
\omega \ell^2(1+2\ln^2[\omega \ell]+c_2 \ln[\omega\ell]+c_3)$. We
note that $\ln[\omega \ell]$ and $\omega \ell\ln^2[\omega \ell]$ are
 signals to indicate the higher-order differential
equations for $A_2$ and $A_3$ in the bulk. Imposing the truncation
on the absorption cross sections, one has $\sigma^{22}_{\rm
abs}=\pi^2 \omega \ell^2$ obtained from scattering the Klein-Gordon
mode off the BTZ black hole. \vspace{1cm}

{\bf Acknowledgments}

TM would like to thank J.-H. Oh and J.-H. Jeong for useful
discussion. This work was supported by the National Research
Foundation of Korea (NRF) grant funded by the Korea government
(MEST) through the Center for Quantum Spacetime (CQUeST) of Sogang
University with grant number 2005-0049409. Y. Myung  was partly
supported by the National Research Foundation of Korea (NRF) grant
funded by the Korea government (MEST) (No.2011-0027293).

\end{document}